\documentclass[12pt,a4paper]{article}
\usepackage{amsfonts,latexsym}
\usepackage{graphicx,color}
\usepackage{dcolumn}
\usepackage{graphicx}
\usepackage{amsmath}
\usepackage{amsfonts}
\usepackage{amssymb}

\renewcommand{\thefootnote}{\fnsymbol{footnote}}

\begin{document}

\vspace{12mm}

\begin{center}
{{{\Large {\bf Superradiant instability in slowly rotating Kerr-Newman black holes  }}}}\\[10mm]

Yun Soo Myung$^{a,b}$\footnote{e-mail address: ysmyung@inje.ac.kr}\\[8mm]

{${}^a$Institute of Basic Sciences and Department  of Computer Simulation,  Inje University Gimhae 50834, Korea\\[0pt]}

{${}^b$Asia Pacific Center for Theoretical Physics, Pohang 37673, Korea}

\end{center}
\vspace{2mm}

\begin{abstract}
We study  the superradiant instability of slowly rotating  Kerr-Newman (sKN) black holes under a charged massive scalar perturbation.
These black holes resemble closer the Reissner-Nordstr\"{o}m  black holes than the Kerr-Newman black holes.
From the scalar  potential analysis, we find that the superradiant instability is  not  allowed in the sKN black holes because the condition for a trapping well is  not compatible with the superradiance condition. However, the rate of energy extraction might  grow exponentially  if the sKN black hole
is placed inside a reflecting cavity. Finally, we obtain  two conditions for the trapping well to possess  quasibound states in  the sKN black holes by analyzing asymptotic scalar potential and far-region wave functions.

\end{abstract}
\vspace{5mm}

\vspace{1.5cm}

\hspace{11.5cm}
\newpage
\renewcommand{\thefootnote}{\arabic{footnote}}
\setcounter{footnote}{0}


\section{Introduction}
Superradiance in the black hole physics is a radiation enhancement process that allows for energy extraction
from the black hole at the classical level~\cite{Brito:2015oca}. It has been  suggested the `rotating black hole-mirror bomb' idea~\cite{Press:1972zz} which states that if the superradiance emerging from a perturbed black hole
were reflected back onto the black hole by a  mirror, an initial perturbation could be made to grow without bound~\cite{Cardoso:2004nk}.
This instability is caused by a reflecting mirror, but the superradiant instability  could occur naturally if a perturbed scalar has a rest mass~\cite{Damour:1976kh}.
On the other hand, if the asymptotic spacetime is de Sitter, it has been shown that  the addition of a mass to the perturbed scalar acts in exactly the opposite way~\cite{Zhu:2014sya,Cardoso:2018nvb,Destounis:2019hca,Mascher:2022pku}.
That is, it depletes  superradiant instabilities that are present when the perturbed scalar is massless.

For a Kerr black hole, if the scalar  has a mass $\mu$, its mass would act as a reflecting mirror.
The superradiant instability depends on two parameters: $a=J/M$  and $M\mu$ ($J$ angular momentum and $M$ mass of the black hole).
The instability gets stronger as $a$ and $\mu$ increase~\cite{Cardoso:2004nk} but  as $a$ decreases with a fixed $\mu$, the unstable modes disappear below a critical value of $a$ because the superradiance condition of $\omega<\omega_c=m \Omega_H$ ($m$ azimuthal number and $\Omega_H$ angular velocity at horizon) violates  for small $a$.
One has found a superradiant instability of the Kerr black hole for $M\mu\gg1$~\cite{Zouros:1979iw}, $M\mu\ll1$~\cite{Detweiler:1980uk}, and   $M\mu\le0.5$~\cite{Dolan:2007mj} for the first corotating mode ($\ell=m=1$).  A scalar potential  including a shape of barrier-well-mirror  is responsible for generating   quasibound states (resonance spectra~\cite{Damour:1976kh} or quasistationary levels~\cite{Dolan:2007mj}) for superradiant instability when satisfying $\omega<\omega_c$ and $\omega<\mu$~\cite{Arvanitaki:2010sy,Konoplya:2011qq}. The later denotes  the bound state condition for getting  asymptotic bound states.
It is worth noting that  the presence of a trapping well is  essential to achieve the superradiant instability
because scalar modes could be localized in the trapping well and  amplified by superradiance to form quasibound states, thus triggering an instability.
If there is no trapping well under a massive scalar wave propagation, the black hole seems to be  superradiantly stable. We note here that  a shortened form of the potential
were employed to sketch the superradiant instability for rotating black branes and strings~\cite{Cardoso:2005vk} because  $\Psi_{\ell m}=rR_{\ell m}$ and a modified tortoise coordinate $z_*$ defined by $dz_*=r^2dr/\Delta$ are used by following Ref.~\cite{Furuhashi:2004jk}.

On the other hand, quasibound states have complex frequencies ($\omega=\omega_{\rm R}+i\omega_{\rm I}$) as the flux passes one way through the outer horizon when solving the Teukolsky equation directly. $\omega_{\rm R}$ represents oscillations and the imaginary part of the frequency
  determines the rate at which the perturbation decays ($\omega_{\rm I}<0$) or grows ($\omega_{\rm I}>0$)  with time.
 For $\omega_{\rm R}=\omega_c$ and $\omega_{\rm R}<\mu$ (stationary resonances), $\omega_{\rm I}$  vanishes and   scalar clouds  for extremal Kerr black holes~\cite{Hod:2012px} and nearly extremal Kerr black holes~\cite{Hod:2013zza} have been found. Making use of  this threshold for superradiant instability has led to Kerr black holes with scalar hair~\cite{Herdeiro:2014goa}. This implies that the hairy Kerr black holes could be found from the growth of  dominant superradiant modes.

Also, the superradiance could represent an amplification  of charged massive scalar waves impinging on a static Reissner-Nordstr\"{o}m (RN) black hole, provided the frequency $\omega$ and the charge $q$ of the scalar wave   obey the superradiance  condition $ \omega<q \Phi_H$ ($\Phi_H$  electric potential at horizon)~\cite{Bekenstein:1973mi}.
Some  aspects of superradiance~\cite{DiMenza:2014vpa} and the absorption cross section of a charged massive scalar~\cite{Benone:2015bst} have been studied in the RN black hole background. We wish to point out  that the studies in the literature~\cite{Furuhashi:2004jk,Hod:2012wmy,Hod:2013nn} have used a shortened potential to show that  contrary to the Kerr case, in the RN case the gravitational attraction between RN black hole and charged massive scalar cannot provide a confinement mechanism  which may  trigger the superradiant instability. At this stage, it is important to note  that the superradiant instability does not arise naturally  from a charged  massive scalar propagation around the RN black holes.
However, the superradiant instability of a charged massive scalar  could be obtained if a cavity is introduced to surround the RN black hole~\cite{Herdeiro:2013pia,Degollado:2013bha,Hod:2013fvl,Hod:2016kpm}.
This is considered  as  the `charged black hole-mirror bomb'. In this case, numerical techniques have used to show the lower bound $q>\mu$ which  considered as  a necessary condition for the superradiant instability~\cite{Herdeiro:2013pia}.

In the Kerr-Newman (KN) black hole background, the superradiant instability condition for a charged massive scalar with $M\mu\leq1$ was firstly obtained as $qQ<\mu M$  which is  regarded as  a  condition for having a trapping well~\cite{Furuhashi:2004jk}. However, this condition is not satisfied when imposing a superradiance condition ($\omega<\omega_c$ with $\omega_c=m\Omega_H+q\Phi_H$) and thus, it may be regarded  as a condition for bound states~\cite{Degollado:2013bha}. Also, we note that  their effective potential $V_{\rm eff}(r)$ is not   a correct form.
Scalar clouds with $\omega=\omega_c$  and $\omega<\mu$ were obtained  in~\cite{Hod:2014baa,Benone:2014ssa} and the absorption cross section of a charged massive scalar was recently computed  to give  a negative cross section for
corotating spherical waves~\cite{Benone:2019all}. Recently, it was reported that the condition for no trapping well is  given by  two  of $qQ>\mu M$ and $r_-/r_+ \le 1/3$~\cite{Xu:2020fgq}. However, their potential based on the analysis is incorrect.
The superradiant stability of a charged massive scalar  based on the correct potential  was discussed in the KN black hole background~\cite{Myung:2022kex}.

We would like to stress that
most black holes are born very slowly rotating~\cite{Fuller:2019sxi}. For example, black holes born from single stars rotate
very slowly for  $a=0.01$ with $M=1$ and  fairly slow rotating black holes born from single stars  are regarded as  those having $a\le0.1$.
But, it holds only at the very first instances of the black hole formation and other studies~\cite{Thorne:1974ve,Novikov:1973kta,Sadowski:2011ka} have demonstrated that accretion can spin up black holes near extremity.
Now, it is curious to introduce  a slowly rotating Kerr-Newman (sKN) black hole.
This black hole could be  obtained  from the KN black hole by confining the first order in $a$ (that is, by taking slow rotation approximation).
We note that the sKN black holes take after more the RN black holes than the KN black holes.
So, it is interesting to investigate superradiant instability of a charged massive scalar propagating around the sKN black holes.
Here, the rotation parameter $a$ is not considered as an important parameter, in comparison with $q$ and $M\mu$ because we confine it to be $a\le 0.1$ for keeping the sKN black holes.
In this work, we wish to study the superradiant instability of the sKN black holes under a charged massive scalar perturbation mainly by analyzing its asymptotic potential and far-region wave function. We obtain two conditions  of a trapping well (\ref{trapping-well}) for getting  quasibound states, whereas the conditions of no trapping well for obtaining bound states are given by (\ref{Notrapping-well}).

\section{Scalar propagation  on the sKN black holes }
Firstly, we introduce  the  KN
black hole expressed in terms of Boyer-Lindquist coordinates
\begin{eqnarray}
ds^2_{\rm KN}&=&-\frac{\Delta}{\rho^2}\Big(dt -a \sin^2\theta d\phi\Big)^2 +\frac{\rho^2}{\Delta} dr^2+
\rho^2d\theta^2 +\frac{\sin^2\theta}{\rho^2}\Big[(r^2+a^2)d\phi -adt\Big]^2 \label{KN}
\end{eqnarray}
with
\begin{eqnarray}
\Delta=r^2-2Mr+a^2+Q^2,~ \rho^2=r^2+a^2 \cos^2\theta,~{\rm and}~a=\frac{J}{M}.
 \label{mps}
\end{eqnarray}
Here, $M$, $Q$, and  $J$ represent the mass, charge, and angular momentum of the KN black hole.
In addition, the electromagnetic potential is given by
\begin{equation}
A_\mu dx^\mu=\frac{Q r}{\rho^2}\Big(-dt,0,0, a\sin^2\theta d\phi\Big). \label{el-back}
\end{equation}
The outer and inner horizons are found  by demanding $\Delta=(r-\tilde{r}_+)(r-\tilde{r}_-)=0$ as
\begin{equation}
\tilde{r}_{\pm}=M\pm \sqrt{M^2-a^2-Q^2}.
\end{equation}

Taking the slow rotation approximation, we  find the slowly rotating  Kerr-Newman (sKN) black hole (so-called Lense-Thirring solution) by keeping up to ${\cal O} (a)$-order~\cite{Lense:1918zz,Hussain:2014cba,Lammerzahl:2018zvb,Hui:2021cpm}
\begin{eqnarray}
ds_{\rm sKN}^2&=&\bar{g}_{\mu\nu}dx^{\mu}dx^{\nu} \nonumber \\
&=& -f_{\rm RN}(r)dt^2+\frac{dr^2}{f_{\rm RN}(r)}
+r^2 (d\theta^2 +\sin^2 \theta d\phi^2)+2a(f_{\rm RN}(r)-1) \sin^2\theta dt d\phi   \label{s-KN}
\end{eqnarray}
with the RN metric function and its electromagnetic potential
\begin{equation}
f_{\rm RN}(r)=1-\frac{2M}{r}+\frac{Q^2}{r^2},~\bar{A}_\mu dx^\mu=\frac{Q}{r}\Big(-dt,0,0, a\sin^2\theta d\phi\Big).
\end{equation}
Clearly,  the line element (\ref{s-KN}) is stationary but non-static because  $dt\to -dt$ changes the
signature of the metric and  it is also axially symmetric (invariance under $d\theta \to -d\theta$).
We note that the sKN spacetime (\ref{s-KN})  inherits the hidden Killing symmetries
of the full solution (\ref{KN}) to ${\cal O} (a)$-order~\cite{Gray:2021toe}.
In this case, the outer and inner horizons are   given by those of the RN black holes as
\begin{equation}
r_\pm=M\pm\sqrt{M^2-Q^2}.
\end{equation}

A charged massive scalar perturbation $\Phi$  on the background of sKN black holes is described  by
\begin{equation}
(\bar{\nabla}^\mu-i q \bar{A}^\mu)(\bar{\nabla}_\mu-i q \bar{A}_\mu)\Phi-\mu^2\Phi=0.\label{phi-eq1}
\end{equation}
Considering  the axis-symmetric
background (\ref{s-KN}), it is convenient to separate the scalar perturbation
into modes
\begin{equation}
\Phi(t,r,\theta,\phi)=\Sigma_{\ell m}e^{-i\omega t+im\phi} P_{\ell}^m
(\theta) R_{\ell m}(r), \label{sep}
\end{equation}
where $P_{\ell}^m(\theta)$ is an associate Legendre polynomial with $-m\le \ell
\le m$ ($Y_{\ell m}(\theta,\phi) \sim P_{\ell}^m(\theta)e^{im\phi}$) and $R_{\ell m}(r)$ satisfies a radial part of the wave
equation. Substituting (\ref{sep}) into (\ref{phi-eq1}), we have
an  associate Legendre equation  for $P_{\ell}^m(\theta)$ and a radial  equation for $R_{\ell m}(r)$ with $\tilde{\Delta}=r^2f_{\rm RN}(r)$~\cite{Hod:2014baa}
\begin{eqnarray}
&& \frac{1}{\sin \theta}\partial_{\theta}\Big(
\sin \theta
\partial_{\theta} P_{\ell}^m(\theta) \Big )+ \left [\ell(\ell+1)-\frac{m^2}{\sin ^2{\theta}} \right ]P_{\ell}^m(\theta) =0,
\label{wave-ang1}
\end{eqnarray}
\begin{eqnarray}
\tilde{\Delta} \partial_r \Big( \tilde{\Delta} \partial_r R_{\ell m}(r) \Big)+U(r)R_{\ell m}(r)=0
\label{wave-rad}
\end{eqnarray}
with
\begin{eqnarray}
U(r)=(\omega r^2-qQr)^2-2am(2M\omega r-qQr-\omega Q^2)-\tilde{\Delta}[\mu^2r^2 +\ell(\ell+1)].
\end{eqnarray}
It is worth noting that Eq. (\ref{wave-rad}) is usually  used  to obtain  exact solutions.

Now, we introduce the tortoise coordinate $r_*$ defined by
\begin{equation}
r_*=\int\frac{dr}{f_{\rm RN}(r)}=r+\frac{r_+^2}{r_+-r_-}\ln(r-r_+)-\frac{r_-^2}{r_+-r_-}\ln(r-r_-)
\end{equation}
 to derive the Schr\"odinger-type equation.
Then,  the radial equation (\ref{wave-rad}) takes a form of
the Schr\"odinger-type equation when setting $\Psi_{\ell m}=r R_{\ell m}$
\begin{equation}
\frac{d^2\Psi_{\ell m}(r_*)}{dr_*^2}+V(r)\Psi_{\ell m}(r_*)=0, \label{sch-eq}
\end{equation}
where  $V(r)$ is found to be~\cite{Benone:2014ssa}
\begin{eqnarray}
V(r)=\Big(\omega-\frac{qQ}{r}\Big)^2 &-&\frac{2am(2M\omega r-qQr-\omega Q^2)}{r^4} \nonumber \\
&-&f_{\rm RN}(r)\Big[\mu^2+\frac{\ell(\ell+1)}{r^2}+\frac{2(Mr-Q^2)}{r^4}\Big].\label{e-pot}
\end{eqnarray}
In the non-rotating limit of $a\to0$, we could recover the scalar potential   from Eq. (\ref{e-pot})
for studying superradiance in the RN black hole~\cite{DiMenza:2014vpa,Benone:2015bst,Herdeiro:2013pia,Degollado:2013bha}.
In  case  of $Q=0$ and $\mu=0$, one finds the scalar potential  from Eq. (\ref{e-pot}) for  investigating the Lense-Thirring black hole~\cite{Vieira:2021ozg}.
In the asymptotic limit, one has $V(r\to \infty)=\omega^2-\mu^2$, while one gets $V(r\to r_+)=(\omega-\omega_c)(\omega-\tilde{\omega}_c)>0$ with $f_{\rm RN}(r_+)=0$ in the near-horizon limit. Here, we have two critical frequencies of  $\omega_c=qQ/r_+$ and $\tilde{\omega}_c=\omega_c+2am/r_+^2>\omega_c$.
Taking the asymptotic limit of Eq. (\ref{sch-eq}) and its near-horizon limit, one has the solutions
\begin{eqnarray}
\Psi&\sim&  e^{-i\sqrt{\omega^2-\mu^2} r_*}(\leftarrow)+{\cal R}e^{+i\sqrt{\omega^2-\mu^2} r_*}(\rightarrow),\quad r_*\to +\infty(r\to \infty) , \label{asymp1}\\
\Psi&\sim& {\cal T} e^{-i\sqrt{(\omega-\omega_c)}\sqrt{(\omega-\tilde{\omega}_c)}}(\leftarrow),\quad r_*\to -\infty(r\to r_+), \label{asymp2}
\end{eqnarray}
where  ${\cal T}({\cal R})$ are the transmission (reflection) amplitudes.

Imposing the flux conservation, we obtain the relation between reflection and transmission coefficients as
\begin{equation}
|{\cal R}|^2=1-\frac{\sqrt{(\omega-\omega_c)}\sqrt{(\omega-\tilde{\omega}_c)}}{\sqrt{\omega^2-\mu^2}}|{\cal T}|^2,
\end{equation}
which  means that only waves with $\omega>\mu$ propagate to infinity and the superradiant scattering  may occur ($\rightarrow,~|{\cal R}|^2>|{\cal I}|^2$) whenever $\omega<\omega_c$ (superradiance condition) is satisfied  because  outgoing waves at the outer horizon reinforce the outgoing waves at infinity.
The absorption cross section is given by
\begin{equation}
\sigma=\sum_{\ell=0}^\infty \sigma_\ell,
\end{equation}
where the partial absorption cross section $\sigma_\ell$ takes the form
\begin{equation}
\sigma_\ell=\frac{\pi}{\omega^2-\mu^2}(2\ell+1)(1-|{\cal R}|^2)= \frac{\pi\sqrt{(\omega-\omega_c)}\sqrt{(\omega-\tilde{\omega}_c)}}{(\omega^2-\mu^2)^{3/2}}(2\ell+1) |{\cal T}|^2.
\end{equation}

On the other hand, one may choose the scalar modes to have an exponentially decay as it tends to zero  at infinity
 \begin{equation}
 {\cal R}_{\ell m} \sim \frac{ e^{-\sqrt{\mu^2-\omega^2} r}}{r} \rightarrow 0 \label{asymp-b}
 \end{equation}
 with the bound state condition of  $\omega<\mu$.

When solving Eq. (\ref{wave-rad}) directly, the frequency $\omega $ is permitted to be complex (small complex modification)
 as~\cite{Dolan:2007mj}
\begin{equation}
\omega=\omega_{\rm R}+i \omega_{\rm I}.
\end{equation}
In this case, the sign of $\omega_{\rm I}$ determines the solution which  is decaying ($\omega_{\rm I}<0$) or growing ($\omega_{\rm I}>0$) in time.
Considering the asymptotic solution form in Eq. (\ref{asymp-b}), the quasibound state condition  can be obtained if Re$[\sqrt{\mu^2-\omega^2}]>0$, tending to zero at infinity, in addition to ingoing at the outer horizon.
We note that  the boundary condition for the well-known quasinormal modes is ingoing at the outer horizon and purely outgoing (and divergent for Re$[\sqrt{\mu^2-\omega^2}]<0$) at infinity. In both quasibound and quasinormal cases~\cite{Percival:2020skc}, imposing a pair of boundary conditions leads to a discrete spectrum of complex frequencies.

Finally, we wish to mention   four cases  for a charged massive  scalar propagating around the sKN black holes based on the potential analysis:\\
Case (i) superradiant scattering: $\omega<\omega_c$ and $\omega>\mu$. \\
Case (ii) superradiant stability: $\omega<\omega_c$ and  $\omega<\mu$  without a positive trapping well. \\
Case (iii) stationary resonances (marginally stable): $\omega=\omega_c$ and $\omega<\mu$.\\
Case (iv) superradiant instability: $\omega<\omega_c$ and  $\omega<\mu$   with  a positive trapping well.\\
Solving the radial equation (\ref{wave-rad}) leads to the real part  of frequency ($\omega_{\rm R}$) and the imaginary part ($\omega_{\rm I}$). In this case, one describes again  the last three cases: \\
Case (ii): $\omega_{\rm I}<0$ and   $\omega_{\rm R}<\omega_c$. The  solution is stable (decaying in time). \\
Case (iii): $\omega_{\rm I}=0$  and $\omega_{\rm R}=\omega_c$. \\
Case (iv): $\omega_{\rm I}>0$ and   $\omega_{\rm R}<\omega_c$. The  solution is unstable (growing in time).

\section{Potential analysis for superradiant instability}
We wish to rewrite Eq. (\ref{sch-eq}) as
\begin{equation}
\frac{d^2\Psi_{\ell m}(r_*)}{dr_*^2}+\Big[\omega^2-V_{\rm sKN}(r)\Big]\Psi_{\ell m}(r_*)=0, \label{nsch-eq}
\end{equation}
where the potential $V_{\rm sKN}(r)$ is given by
\begin{eqnarray}
V_{\rm sKN}(r)&=& \mu^2-\frac{2(M\mu^2-qQ\omega)}{r} +\frac{Q^2(\mu^2-q^2)+\ell(\ell+1)}{r^2}\nonumber \\
&+&\frac{2M-2M\ell(\ell+1)+2am(2M\omega-qQ)}{r^3}\nonumber \\
    &-&\frac{4M^2+2Q^2-Q^2\ell(\ell+1)+2amQ^2 \omega}{r^4}+\frac{6M Q^2}{r^5}-\frac{2Q^4}{r^6}. \label{new-pot}
\end{eqnarray}
An  asymptotic  form of the potential is given by
\begin{equation}
V_{\rm a}(r)=\mu^2-\frac{2(M\mu^2-qQ\omega)}{r} \label{asym-p}
\end{equation}
which appears  in the asymptotic region. This potential could be used to find the condition for a trapping well as
\begin{equation}
V'_{\rm a}(r)>0 \quad  \to M\mu^2> qQ\omega, \label{cond-t}
\end{equation}
while the condition for no trapping well is realized as
\begin{equation}
V'_{\rm a}(r)<0 \quad\to M\mu^2< qQ\omega. \label{nocond-t}
\end{equation}
However, Eq.(\ref{cond-t}) [Eq.(\ref{nocond-t})] is not a sufficient condition for a trapping well [no trapping well].
We have to find the other conditions. For this purpose,
we introduce  the far-region potential appeared in the large $r$ region
\begin{equation}
V_{\rm fr}(r)=\mu^2-\frac{2(M\mu^2-qQ\omega)}{r}+\frac{Q^2(\mu^2-q^2)+\ell(\ell+1)}{r^2}, \label{fr-p}
\end{equation}
where the last term plays a crucial role of making a trapping well.

In the non-rotating limit of $a\to 0$, one finds the potential for a charged massive scalar propagating  around the RN black holes~\cite{DiMenza:2014vpa,Benone:2015bst,Herdeiro:2013pia,Degollado:2013bha} as
\begin{eqnarray}
V_{\rm RN}(r)&=& \mu^2-\frac{2(M\mu^2-qQ\omega)}{r}+\frac{Q^2(\mu^2-q^2)+\ell(\ell+1)}{r^2} \nonumber \\
&+&\frac{2M-2M\ell(\ell+1)}{r^3}-\frac{4M^2+2Q^2-Q^2\ell(\ell+1)}{r^4}+\frac{6M Q^2}{r^5}-\frac{2Q^4}{r^6} \label{RN-pot}
\end{eqnarray}
whose asymptotic  and far-region potentials are still given by   Eqs. (\ref{asym-p}) and (\ref{fr-p}).
\begin{figure*}[t!]
   \centering
  \includegraphics{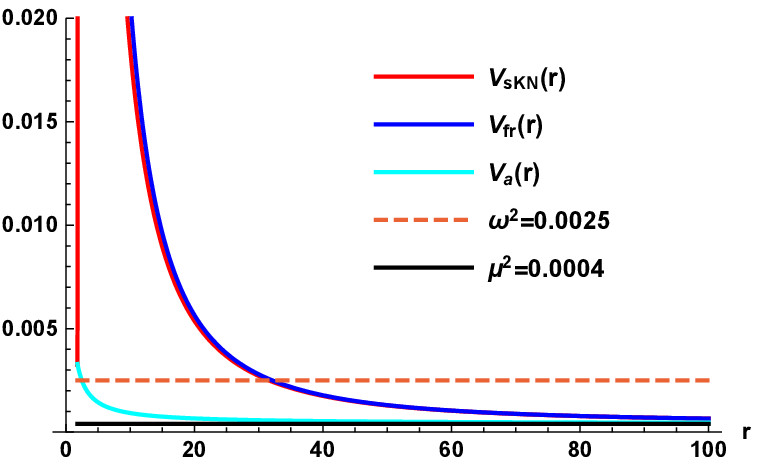}
   \hfill%
  \includegraphics{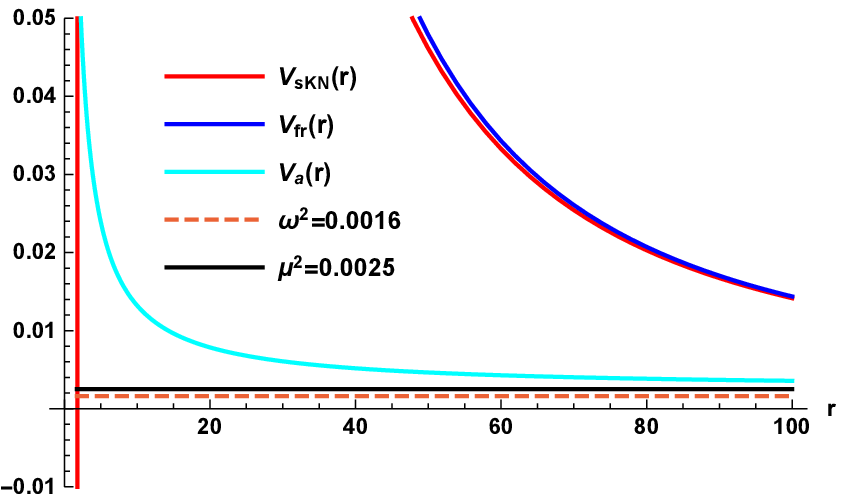}
\caption{(Left) Superradiant scattering  potential $V_{\rm sKN}(r)$   as function of $r\in[r_+ =1.8,100]$ with $M=1,Q=0.6,\omega=0.05,a=0.1,m=1,q=0.1,\ell=1,\mu=0.02$.
The height of potential barrier is $0.114 (\gg \omega^2)$ at $r=2.64$.
 (Right) Superradiant stable potential $V_{\rm sKN}(r)$ as function of $r\in[r_+=1.7,100]$ with $M=1,Q=0.7,\omega=0.04,a=0.1,m=10,q=2,\ell=10,\mu=0.05$. The height of barrier is $4.6 (\gg \omega^2)$ at $r=2.72$. We check the conditions of $\omega<\mu$ and $\omega<\omega_c=0.8$ to have a superradiant stability with $V'_{\rm a}(r)<0$.  }
\end{figure*}
\begin{figure*}[t!]
   \centering
  \includegraphics{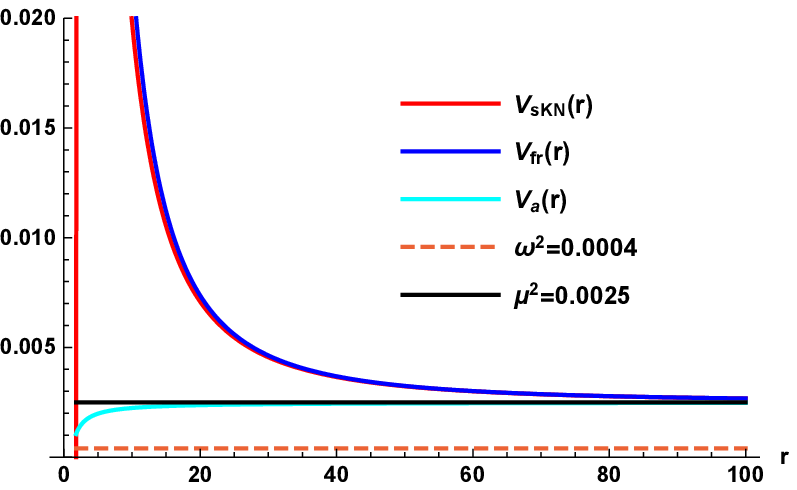}
   \hfill%
  \includegraphics{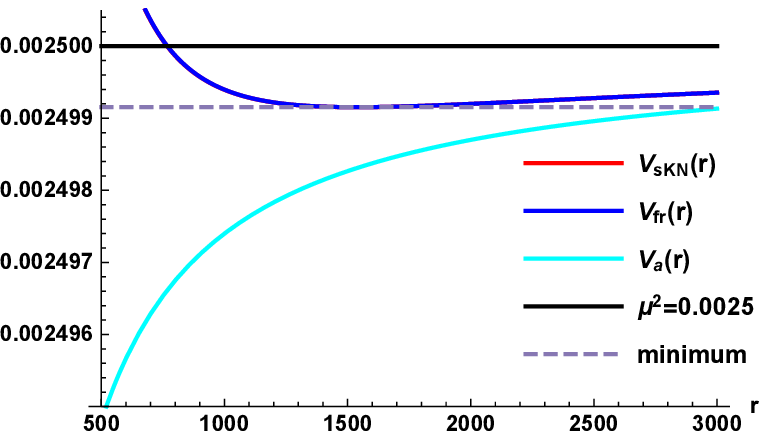}
\caption{(Left) Superradiant stable  potential $V_{\rm sKN}(r)$   as function of $r\in[r_+ =1.8,100]$ with $M=1,Q=0.6,\omega=0.02,a=0.1,m=1,q=0.1,\ell=1,\mu=0.05$.
 The height of barrier is $0.11 (\gg \omega^2)$ at $r=2.66$. We check the conditions of $\omega<\mu$ and $\omega<\omega_c=0.03$ to have a superradiant stability, but $V'_{\rm a}(r)>0$ implies a trapping well. (Right) Asymptotic forms of  $V_{\rm sKN}(r)\simeq V_{\rm fr}(r)$ indicate a tiny well located at $r=1535$. $V_{\rm a}(r)$ approaches them for $r>1535$. }
\end{figure*}

At this stage, it is worth mentioning that $V_{\rm sKN}(r)$ and $ V_{\rm RN}(r)$ are $\omega$-dependent potentials, compared to the standard potentials appeared in Schr\"{o}dinger-like problems. However, some information on the superradiant instability could be extracted  by investigating these potentials.

\begin{figure*}[t!]
   \centering
  \includegraphics{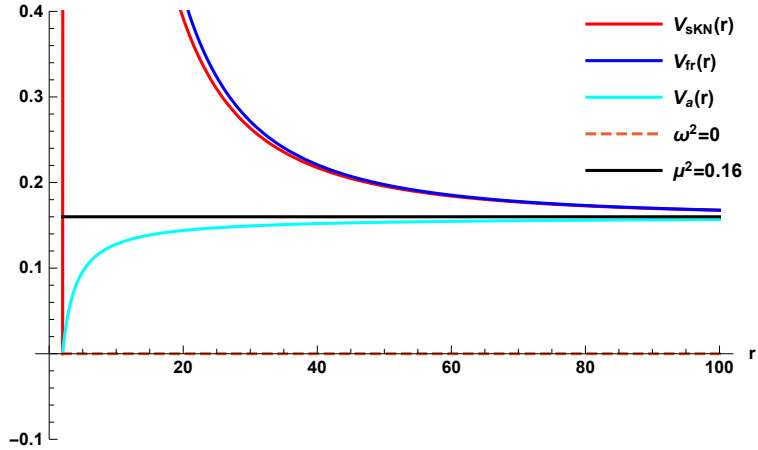}
\caption{Stationary  resonances   potential $V_{\rm sKN}(r)$   as function of $r\in[r_+ =1.999,100]$ with $M=1,Q=0.01,\omega=0.001,a=0.1,m=10,q=0.2,\ell=10,\mu=0.4$.
We have $V_{\rm sKN}(r_+)=0\simeq \omega^2$ and
the height of potential barrier is $4.14 (\gg \omega^2)$ at $r=3.01$.
We note   $\omega=\omega_c=0.001$ and $\omega<\mu$ to meet the condition for obtaining scalar clouds.   }
\end{figure*}
Hereafter, we choose $M=1$ such that $M\mu$ becomes $\mu$  for a simple analysis.
First of all,
 we consider the superradiant scattering [Case (i)]. We display the corresponding potential in (Left) Fig. 1, indicating that
 $\mu,~\omega,~q\Phi_H \ll1$  and $\mu <\omega_c<\omega$ with $a=0.1$ to give the negative absorption cross section.
 In addition, the superradiant stability [Case (ii)] could be achieved for $\omega<\mu$, $\omega<\omega_c$, and $q>\mu$ as is shown in (Right) Fig. 1. It includes no  trapping well,  being consistent with $V'_{\rm a}(r)<0$.

It is curious to note that (Left) Fig. 2 corresponds to a  superradiantly stable potential  because we could not find a trapping well  for $\omega<\omega_c=0.03$ and $\omega<\mu$.
In this case, however, we observe   $V'_{\rm a}(r)>0$ which may imply the superradiant instability. So,  $V'_{\rm a}(r)>0$ contradicts to our expectation of no trapping well.
We wish to resolve it. We find from (Right) Fig. 2 that a tiny  well is located at a very large distance of $r=1535$ in $V_{\rm aKN}(r)\simeq V_{\rm fr}(r)$, but it does not affect the superadiant stability. It indicates that $V'_{\rm a}(r)>0$ implies either a trapping well or a tiny well. Hence, one has to find the other condition for a trapping well in the next section.

To visualize stationary resonances [Case (iii)], we observe the corresponding potential $V_{\rm sKN}(r)$ with $a=0.1$ and $\mu=0.4$ in Fig. 3.
This  is similar to (Left) Fig. 2, showing the superradiant stability except $\omega=\omega_c$. But, the effect of  imposing $\omega=\omega_c$ appears in the near-horizon region.

\begin{figure*}[t!]
   \centering
  \includegraphics{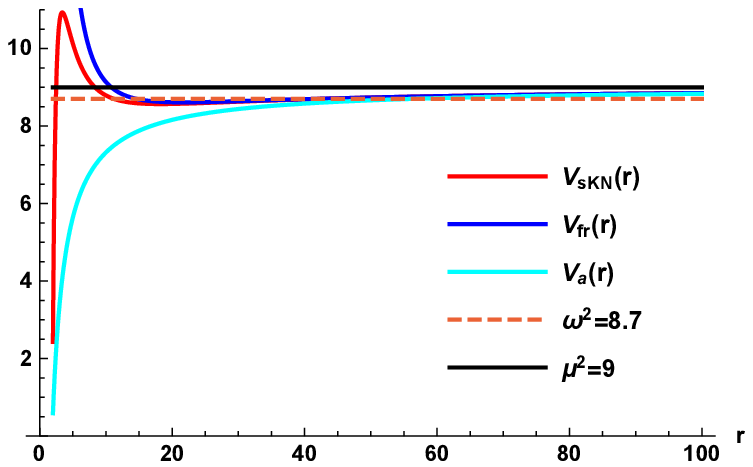}
  \hfill%
  \includegraphics{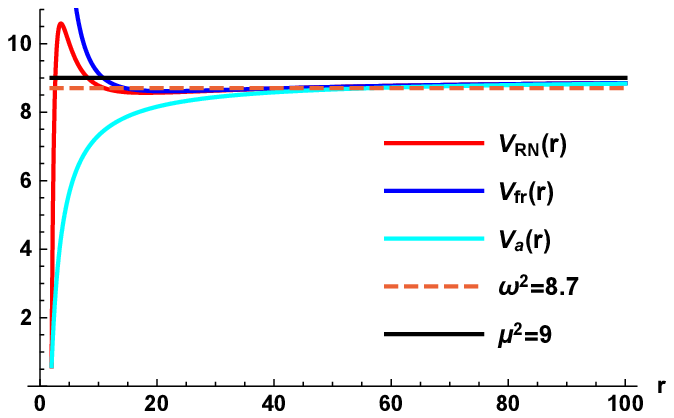}
\caption{(Left) Superradiant unstable  potential $V_{\rm sKN}(r)$   as function of $r\in[r_+ =1.999,100]$ with $M=1,Q=0.01,\omega=2.95,a=0.1,m=13,q=20,\ell=13,\mu=3$.
The potential at $r=r_+$ is 2.43 and
a peak  of potential is $10.93(> \omega^2)$ at $r=3.41$.  A trapping well (local minimum) is located at $r=18.24$.
We note  $\omega<\mu$, but $\omega>\omega_c=0.1$ fails to satisfy the superradiance condition ($\omega<\omega_c$). (Right) Superradiant unstable  potential $V_{\rm RN}(r)$ with $a=0$. One has $V_{\rm RN}(r_+)=0.58$ and
a peak of potential is $10.59(> \omega^2)$ at $r=3.59$.  A trapping well is located at $r=18.06$. With  $\omega<\mu$, we note that $\omega>\omega_c=q\Phi_H=0.1$ fails to satisfy the superradiance condition.  }
\end{figure*}

As a specific example for Case (iv), we wish to introduce  a sKN  potential [(Left) Fig. 4) with $a=0.1$ and $q>\mu$.
It seems  that  quasibound states of a charged massive scalar  do not contain superradiant states.
That is, the condition for a trapping well  and the superradiance  condition  ($\omega<\omega_c$) cannot be satisfied simultaneously   because of $\omega>\omega_c$. This implies that the superradiant instability is not found from a charged  massive scalar propagating  around the sKN black holes.
In this case, the scalar mass $\mu$ is no longer a reflecting mirror. The same feature is found from the RN potential [(Right) Fig. 4]~\cite{Herdeiro:2013pia}.

However, there might be a way to obtain  the superradiant instablity in the sKN black hole background.
If a mirror (cavity) is placed  at some radial coordinate $r_m$ outside the outer horizon, the asymptotic boundary condition is modified such that the scalar mode
vanishes exactly at $r=r_m~[R_{\ell m}(r_m)=0]$  instead of Eq. (\ref{asymp-b}) and its proper frequency may be  determined by imposing  the position of the mirror.
The scalar mode might have frequencies that  are in the superradiant regime ($\omega_{\rm R}<\omega_c$) because  one can place the mirror  at arbitrarily close to the black hole horizon
($r_m\to r_+$ in the $qQ\to \infty$  limit).
In the non-rotating limit of $a\to 0$ (the RN black hole)~\cite{Herdeiro:2013pia,Degollado:2013bha}, one has found the real and  imaginary part of the resonance frequency as~\cite{Hod:2013fvl}
\begin{equation}
\omega_{\rm R}=\omega_c(1-x_m),\quad \omega_{\rm I}=\omega_c  \sqrt{\frac{x_m^3}{\tau}},
\end{equation}
where
\begin{equation}
\omega_c= \frac{qQ}{r_+},~x_m=\frac{r_m-r_+}{r_+},~\tau=\frac{r_+-r_-}{r_+}
\end{equation}
in the asymptotic regime of $qQ\gg \tau/x_m\gg1$. We note here that $\omega_{\rm R}<\omega_c$ and  $\omega_{\rm I}>0$ implies a superradiant  instability of the charged black hole-mirror system.

\section{Far-region and asymptotic wave functions}
It is crucial to find  the scalar wave forms in the  far-region  to distinguish between quasibound states (trapping well)  and  bound states (no trapping well).
This is because the condition of $V'_{\rm a}(r)>0$ in Eq. (\ref{cond-t}) is not a sufficient condition for getting a trapping well.

In the far-region where we may take $r_*\simeq r$, we obtain an  equation from  (\ref{nsch-eq}) together with (\ref{fr-p}) as
\begin{equation}
\Big[\frac{d^2}{dr^2}+\omega^2-V_{\rm fr}(r)\Big]\Psi_{lm}(r)=0
\end{equation}
whose solution is given exactly by the confluent  Hypergeometric function $U(a,b;cr)$ as
\begin{eqnarray}
\Psi_{\ell m}(r)&=&c_1 e^{-\sqrt{\mu^2-\omega^2}r} \Big(2\sqrt{\mu^2-\omega^2} r\Big)^{\frac{1}{2}+k} \nonumber \\
 &\times & U\Big(\frac{1+2k}{2}-\frac{M\mu^2-qQ\omega}{\sqrt{\mu^2-\omega^2}},1+2k;2\sqrt{\mu^2-\omega^2} r\Big) \label{wavef-1}
\end{eqnarray}
with
\begin{equation}
k=\frac{1}{2}\sqrt{1+4[\ell(\ell+1)+Q^2(\mu^2-q^2)]}.
\end{equation}
 Here,  we observe a bound state of  $e^{-\sqrt{\mu^2-\omega^2}r}$ with $\omega<\mu$ appeared  in (\ref{asymp-b}).
 \begin{figure*}[t!]
   \centering
  \includegraphics{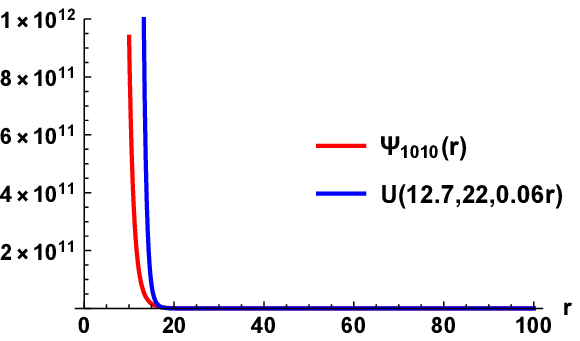}
  \hfill%
  \includegraphics{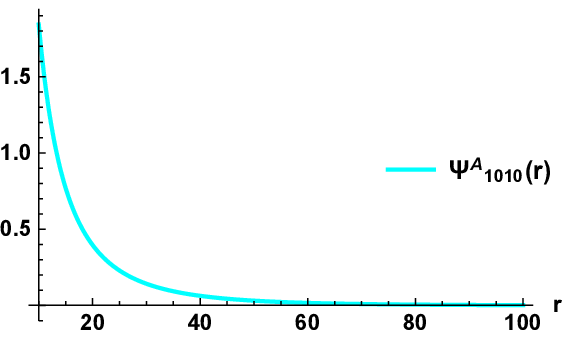}
\caption{ (Left) Bound state function $\Psi_{1010}(r)$ and its confluent hypergeometric function $U(12.7,22;0.06r)$ as $r\in[10,100]$ without trapping well. (Right) Its asymptotic wave function $\Psi^{\rm A}_{1010}(r)$ represents an asymptotic  bound state.}
\end{figure*}
 \begin{figure*}[t!]
   \centering
  \includegraphics{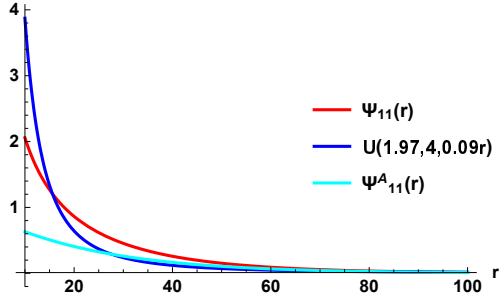}
\caption{ Bound state function $\Psi_{11}(r)$, its confluent hypergeometric function $U(1.97,4;0.09r)$, and its asymptotic wave function $\Psi^{\rm A}_{11}(r)$ as $r\in[10,100]$ without trapping well.}
\end{figure*}
\begin{figure*}[t!]
   \centering
  \includegraphics{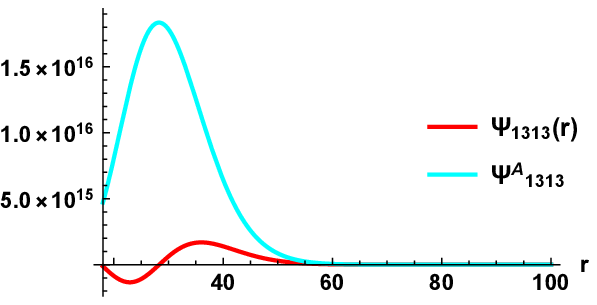}
  \hfill%
  \includegraphics{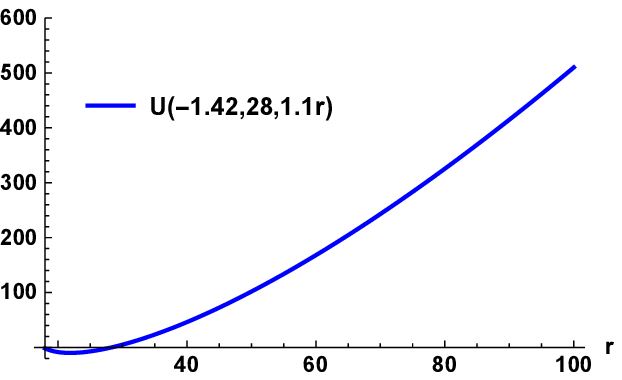}
\caption{(Left) Quasi-bound state of $\Psi_{1313}(r)$ as function of $r\in[22,100]$ with trapping well. Here, we start with $r=18$ because a local minimum of $V_{\rm sKN}(r)$ in Eq. (\ref{new-pot})  is located at $r=18.2$. (Right) Confluent hypergeometric function $U(-1.42,28;1.1r)$ represents  an increasing function of $r$ approximately.}
\end{figure*}
Furthermore, we find  some information from the large $r$-form of $U(a,b;cr)$ as
\begin{equation}
U(a,b;cr\to \infty)\rightarrow\quad  \frac{1}{(cr)^{a}}\Big[1-\frac{a(1+a-b)}{cr} +{\cal O}\Big(\frac{1}{cr}\Big)^2\Big] \label{large-U}
\end{equation}
which implies approximately that one finds  a decreasing function $U(a,b;cr)$ for a positive $a$, whereas one has an increasing function  for a negative $a$.
Substituting  Eq. (\ref{large-U}) into Eq. (\ref{wavef-1}) leads to the asymptotic wave function
as
\begin{equation}
\Psi^{\rm A}_{\ell m}(r)\simeq e^{-\sqrt{\mu^2-\omega^2}r} \Big(2\sqrt{\mu^2-\omega^2} r\Big)^{\frac{M\mu^2-qQ\omega}{\sqrt{\mu^2-\omega^2}}}. \label{wavef-2}
\end{equation}

We consider three cases with $q>\mu$ only.
Considering the potential  without trapping well whose asymptotic derivative is negative ($V_{\rm a}'(r)<0$) [(Right) Fig. 1], $\Psi_{1010}(r)$ in Fig. 5 shows an exponentially decaying mode (bound state). Also, we have a rapidly decreasing function  $U(12.7,22;0.06r)$ and the asymptotic wave function $\Psi^{\rm A}_{1010}(r)$ is an exponentially decreasing function. This case represents no trapping well clearly.

We introduce  an interesting  potential without apparently trapping well whose asymptotic derivative is positive ($V_{\rm a}'(r)>0$) [(Left) Fig. 2].  Its wave function $\Psi_{11}(r)$ and its confluent hypergeometric function $U(1.97,4;0.1r)$ indicate monotonically decreasing modes [Fig. 6]. Also, its asymptotic wave function $\Psi^{\rm A}_{11}(r)$ is a slowly decreasing function. Although this potential includes a tiny well located at $r=1535$ [see (Right) Fig. 2], it could be  neglected effectively. After analyzing far-region wave function, we conclude that it does not include  an apparently trapping well.

Let us observe a radial mode  $\Psi_{\ell m}(r)$ for a trapping well [see (Left) Fig. 4].
As is shown in (Left) Fig.7, Eqs.(\ref{wavef-1}) and (\ref{wavef-2}) show   quasi-bound states.
In this case, one has an increasing function $U(-1.42,28;1.1r)$ appeared in (Right) Fig. 7.

Therefore, the quasibound state  could be achieved when the first argument of $U(a,b;cr)$ is negative as
\begin{equation}
a<0 \to \quad \frac{M\mu^2-qQ\omega}{\sqrt{\mu^2-\omega^2}}>k+\frac{1}{2} \label{trap-well}
\end{equation}
which is considered as  the other condition for  trapping well.
 On the other hand,
the bound state   could be found  when the first argument of $U(a,b;x)$ is positive as
\begin{equation}
a>0 \to \quad \frac{M\mu^2-qQ\omega}{\sqrt{\mu^2-\omega^2}}<k+\frac{1}{2},\label{no-tw}
\end{equation}
which is regarded as the other condition for no trapping well.

At this stage, we have the same condition (\ref{trap-well}) for having  a trapping well under a charged massive scalar propagating   around the RN black holes [see (Right) Fig. 5] because its  asymptotic and far-region  potentials is given exactly  by and (\ref{asym-p}) and  (\ref{fr-p}), respectively. Also, no trapping well is  allowed under the condition of (\ref{no-tw}).

Finally, we obtain   two conditions for getting a trapping well as
\begin{equation}
M\mu^2> qQ\omega~[V'_{\rm a}(r)>0]~{\rm and}~\frac{M\mu^2-qQ\omega}{\sqrt{\mu^2-\omega^2}}>k+\frac{1}{2}. \label{trapping-well}
\end{equation}
On the other hand, two conditions for no trapping well are given by
\begin{equation}
M\mu^2< qQ\omega~[V'_{\rm a}(r)<0]~{\rm and}~\frac{M\mu^2-qQ\omega}{\sqrt{\mu^2-\omega^2}}<k+\frac{1}{2}, \label{Notrapping-well}
\end{equation}
where the former condition could be written as
\begin{equation}
\frac{M\mu}{qQ} <\frac{\omega}{\mu} <1,
\end{equation}
which may imply one condition for the  superradiant stability as $qQ>M\mu$ appeared in Ref.\cite{Xu:2020fgq}.
The other case of $M\mu^2> qQ\omega~[V'_{\rm a}(r)>0]$ and $\frac{M\mu^2-qQ\omega}{\sqrt{\mu^2-\omega^2}}<k+\frac{1}{2}$ represents a tiny well located at a very large distance ($r=1535$), which is considered as no trapping well apparently. This explains why one needs to have two conditions to specify a trapping well.

\section{Discussions}

We have studied  the superradiant instability of slowly rotating  Kerr-Newman (sKN) black holes under a charged massive scalar perturbation.
It is worth noting that these black holes take after more the RN black holes than the KN black holes because
 the rotation parameter restricting  $a\le 0.1$ is not  an important one, in comparison with $q$ and $\mu$.

For $q>\mu$, we have obtained a potential  with a  trapping well.
However, the superradiance condition of  $\omega<\omega_c$  fails to be satisfied, being similar to the RN black holes.
If a mirror (cavity) is introduced at some radius $r_m$  outside the outer horizon, the superradiance condition could be managed to be satisfied.

Importantly, we obtain two conditions  of a trapping well (\ref{trapping-well}) for getting  quasibound states. On the other hand,  two conditions of no trapping well for obtaining bound states are given by (\ref{Notrapping-well}). Furthermore,  we could apply these conditions to a charged massive scalar propagating around the RN black holes, too.

Finally, it is worth noting that    the superradiant instability  could be achieved if  all of two conditions for a trapping well (\ref{trapping-well}), superradiance condition ($\omega<\omega_c$),
and bound state condition ($\omega<\mu$) are satisfied. Also, the superradiant stability is found when all of  two conditions for no trapping-well (\ref{Notrapping-well}), superradiance condition $\omega<\omega_c$,
and bound state condition ($\omega<\mu$) are satisfied.

 \vspace{0.5cm}

{\bf Acknowledgments}
 \vspace{0.5cm}

This work was supported by a grant from Inje University for the Research in 2021 (20210040).

\end{document}